\documentclass[prl,twocolumn,showpacs,aps]{revtex4}
\usepackage{graphics}
\usepackage{dcolumn}
\usepackage[sort&compress]{natbib}
\newcommand{\ab}{{\bf a}}
\newcommand{\bb}{{\bf b}}
\newcommand{\cb}{{\bf c}}
\newcommand{\db}{{\bf d}}
\newcommand{\eb}{{\bf e}}
\newcommand{\fb}{{\bf f}}
\newcommand{\gb}{{\bf g}}
\newcommand{\hb}{{\bf h}}

\newcommand{\nb}{{\bf n}}
\newcommand{\rb}{{\bf r}}
\newcommand{\tb}{{\bf t}}
\newcommand{\abt}{{\bf\widetilde{a}}}
\newcommand{\nbt}{{\bf\widetilde{n}}}

\newcommand{\Ri}{{{\bf R}_i}}
\newcommand{\Lb}{{\bf L}}
\newcommand{\Mb}{{\bf M}}

\newcommand{\Ec}{{\cal E}}
\newcommand{\mOmega}{{\mit\Omega}}
\newcommand{\sump}{\mathop{{\sum}'}}
\begin{document}
\title{Spherical spin-orientation degeneracy of basic antiferromagnetic
configurations\\
due to the dipolar interaction in cubic lattices}
\author{Eugene V. Kholopov}
\email{kholopov@casper.che.nsk.su}
\affiliation{Institute of Inorganic Chemistry of the Siberian Branch of
the Russian Academy of Sciences, 630090 Novosibirsk, Russia}
\begin{abstract}
Based on a simple general relation for the Lorentz field, the precise
values are obtained for the energies of ferromagnetic and basic
antiferromagnetic states in sc, bcc, fcc, and diamond cubic lattices.
Within both the 'nearest-neighbor' and group-theoretic approaches, a large
variety of antiferromagnetic states revealed as corresponding to the
lowest energies is addressed, with recognizing their spherical
spin-orientation degeneracy, which turns out to be rather special in fcc
lattices.
\end{abstract}
\maketitle

The dipolar interaction of spins is typical of dilute
magnets \cite{Vlec37} and so remains actual upon describing modern 
complex compounds where the direct exchange is either depressed
\cite{Niem72,Holm75,Rose90,Fern00} of frustrated \cite{Palm00}. In this
connection, dipolar solutions in cubic structures are interesting
themselves \cite{Holm75,Rose90,Niem73,Mass87} and also enable one to
understand magnetic ordering in structures of a lower symmetry 
\cite{Fern00,Coh55b,Fuji87,Stei98}. It is known that antiferromagnetic 
ground states are typical of the sc \cite{Saue40} and diamond 
\cite{Whit93} structures, whereas the bcc and fcc lattices are 
ferromagnetic \cite{Rose90,Saue40,Lutt46} in the thermodynamic limit 
\cite{Fern00,Whit93,Grif68,Khol04}.

In order to describe basic spin configurations, the 'nearest-neighbor'
\cite{Saue40} and group-theoretic \cite{Lutt46} treatments are usually
applied \cite{Fern00}. However, an explicit motif of spin arrangement in
shortest-distance directions \cite{Saue40}, that is characteristic of the
dipolar interaction, is not still recognized in the bcc lattice.
Nevertheless, the same motif gives rise to the widespread standpoint that
basic antiferromagnetic arrangements are attached to crystallographic axes
\cite{Fern00,Whit93,Bouc93}. This is at odds with spherical degeneracy of
ferromagnetic configurations \cite{Rose90}, as well as with the fact that 
the spherical degeneracy of antiferromagnetic states in sc lattices has 
already been proved \cite{Lutt46}. This is the reason that here we study 
this question addressed to all cubic lattices in a systematic manner, with 
treating spins as classical vectors.

As for the problem of summation of dipolar lattice series that is rather
tedious conventionally \cite{Fern00,Brue70,Colp71}, here we propose a
simple general approach based on introducing fictitious charges
\cite{Evje32}. Let us consider the general case of a triclinic crystal
with lattice constants $a$, $b$, and $c$ along non-orthogonal triclinic
directions $\eb_a$, $\eb_b$, and $\eb_c$, respectively, and with $n$
moments $\Mb_j$ localized at the positions $\tb_j$ in a unit cell,
providing that $\Mb_j$ can differ from one other. Concentrating on a 
$j$th sublattice built up of parallel moments $\Mb_j$ at a given $j$, 
it is expedient to cast $\Mb_j$ in terms of triclinic coordinates
\begin{equation}\label{Eq1}
\Mb_j=M_{ja}\eb_a+M_{jb}\eb_b+M_{jc}\eb_c\equiv[M_{ja},M_{jb},M_{jc}] ,
\end{equation}
where the square brackets are used in order to distinguish this
representation from a Cartesian one. Then the unit cell of this
sublattice, with $\Mb_j$ at its center, can be modified by three point
charge species
\begin{equation}\label{Eq2}
q_{j1}=-\frac{M_{ja}}{a} ,\quad q_{j2}=-\frac{M_{jb}}{b} ,\quad
q_{j3}=-\frac{M_{jc}}{c}
\end{equation}
at the unit-cell positions
\begin{equation}\label{Eq3}
\rb_1=[{\textstyle\frac{1}{2}}a,0,0] ,\quad
\rb_2=[0,{\textstyle\frac{1}{2}}b,0] ,\quad
\rb_3=[0,0,{\textstyle\frac{1}{2}}c] ,
\end{equation}
but the charges $-q_{j1}$, $-q_{j2}$, and $-q_{j3}$ are assumed at
$-\rb_1$, $-\rb_2$, and $-\rb_3$, respectively. As a result, one can see
that the total dipolar moment of the unit cell is zero and all the charges
above cancel one other upon summing over the lattice. Therefore, the field
exerted by this modified sublattice on any moment $\Mb$ at the position
$\tb$ is determined by the convergent lattice series, which can be easily
obtained from the appropriate expansion of a Coulomb lattice series, as
will be discussed elsewhere \cite{Khol01}. This is a generalized Lorentz
field of the form
\begin{eqnarray}\label{Eq4}
&&\Lb_j(\tb)=\sump_i\Bigl\{\Bigl[\frac{3\bigl((\Ri-\tb)\Mb_j\bigr)(\Ri
-\tb)}{|\Ri-\tb|^5}-\frac{\Mb_j}{|\Ri-\tb|^3}\Bigr]\nonumber\\
&&\quad{}+\sum_mq_{jm}\Bigl[\frac{\Ri-\rb_m-\tb}{|\Ri-\rb_m
-\tb|^3}-\frac{\Ri+\rb_m-\tb}{|\Ri+\rb_m-\tb|^3}\Bigr]\Bigr\} ,
\end{eqnarray}
where the summation is over the $j$th sublattice described by $\Ri$, the
prime on the summation sign implies missing the cases of zero
denominators, $\bigl((\Ri-\tb)\Mb_j\bigr)$ stands for the scalar product.
The first term in the curly braces in (\ref{Eq4}) is the direct sum of dipole
contributions and the second one is the Lorentz field itself that is
responsible for the topological boundary effects \cite{Khol04}. Note that
the convergence of (\ref{Eq4}) takes place if the summation over $i$ is
the outer procedure. Keeping (\ref{Eq4}) in mind and taking all the
sublattices into account, the total bulk dipolar energy per unit cell is
as follows:
\begin{equation}\label{Eq5}
\Ec=-\frac{1}{2}\sum_{j_1,j_2}\bigl(\Mb_{j_1}\Lb_{j_2}
(\tb_{j_1}-\tb_{j_2})\bigr) .
\end{equation}

Here we focus on the four cubic structures with the lattice spacing $a$:
the sc, bcc, fcc, and diamond ones. The volume $v_s$ per spin is equal to
$a^3$, $a^3/2$, $a^3/4$, and $a^3/8$, respectively. All the spin values
$|\Mb|$ are supposed to be the same, so that the energy per spin
$\Ec_s=\Ec/n$ is of interest. The energies for ferromagnetic spin
arrangements calculated from (\ref{Eq2})--(\ref{Eq5}) are shown in
Table \ref{Table1} and agree with the fact that the direct dipole
\begin{table}[t]
\caption{The bulk dipolar energy $\Ec_s$ per spin in units of 
$\Mb^2/v_s$ for the ferro- (F) and basic antiferromagnetic (AF) 
states in cubic structures, with indicating the three- (3D) and 
two-dimensional (2D) degeneracy in the latter case.}\label{Table1}
\begin{ruledtabular}
\begin{tabular}{lllcc}
\raisebox{-1.3ex}[0pt][0pt]{Structure}&
\raisebox{-1.3ex}[0pt][0pt]{\hspace{1em}
$\Ec_s$ (F)\tablenotemark[1]}&
\raisebox{-1.3ex}[0pt][0pt]{\hspace{0.6em}$\Ec_s$ (AF)}&
\multicolumn{2}{c}{AF degeneracy}\\
\cline{4-5}
&&&3D$\lefteqn{\tablenotemark[1]}$&2D$\lefteqn{\tablenotemark[2]}$\\
\hline
sc&$-2.094395$\tablenotemark[3]&$-2.676789$&
1$\lefteqn{\tablenotemark[4]}$&---\\
bcc&$-2.094395$\tablenotemark[3]&$-1.985921$\tablenotemark[5]&
2&3$\lefteqn{\tablenotemark[4]}$\\
fcc&$-2.094395$\tablenotemark[3]&$-1.807574$&
4$\lefteqn{\tablenotemark[6]}$&3\tablenotemark[4]\tablenotemark[7]$\!
{}+3$\tablenotemark[8]$\:$\\
diamond&$-2.094395$\tablenotemark[3]&$-3.196851$&
2&3$\lefteqn{\tablenotemark[4]}$
\end{tabular}
\end{ruledtabular}
\tablenotetext[1]{The direction of a reference spin species is arbitrary.}
\tablenotetext[2]{Reference spins lie in one of the three basal planes.}
\tablenotetext[3]{This is just the value of $-2\pi/3$, in accord with
(\ref{Eq6}) \cite{Lutt46}.}
\tablenotetext[4]{Here the cases are contained of collinear spins inherent
in the sc \cite{Saue40}, bcc \cite{Khol01}, fcc \cite{Saue40}, and diamond
\cite{Whit93} lattices.}
\tablenotetext[5]{This value was discussed earlier \cite{Lutt46} as 
associated with spins along different crystallographic axes in the two sc 
sublattices.} 
\tablenotetext[6]{Two-parameter sets of states at a given reference spin.}
\tablenotetext[7]{One-parameter continuous sets of states.}
\tablenotetext[8]{One-parameter piecewise continuous sets of states.}
\end{table}
contribution to (\ref{Eq4}) upon summation over cubic shells is zero by
symmetry, so that the bulk dipolar energy per spin is reduced to the
contribution of the Lorentz field itself \cite{Lore16}:
\begin{equation}\label{Eq6}
\Ec_s=-\frac{2\pi\Mb^2}{3v_s} .
\end{equation}

\begin{figure}[t]
\resizebox{0.95\hsize}{!}{\includegraphics{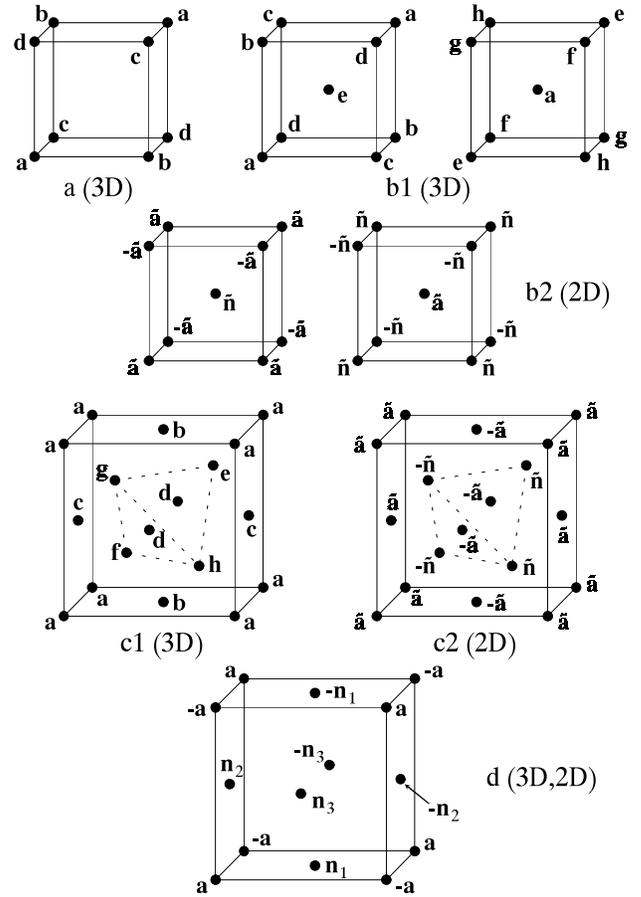}}
\caption{3D basic antiferromagnetic configurations in the sc (a), bcc
(b1), and diamond (c1) structures, with spins specified by
(\protect{\ref{Eq9}}) and (\protect{\ref{Eq12}}). 2D configurations
typical of the bcc (b2) and diamond (c2) structures contain spins driven
by (\protect{\ref{Eq14}}). A basic antiferromagnetic fcc structure (d) is
common to 3D and 2D configurations described by (\protect{\ref{Eq15}}) and
either by (\protect{\ref{Eq20}})--(\protect{\ref{Eq23}}) or by
(\protect{\ref{Eq28}})--(\protect{\ref{Eq30}}) as an example,
respectively.}\label{Fig1}
\end{figure}

This approach is also fruitful upon calculating basic antiferromagnetic
states, which can be decomposed into sc ferromagnetic sublattices. The
results are exhibited in Table \ref{Table1} as well. To address a large
variety of such states revealed, we resort to the description of Luttinger
and Tisza \cite{Lutt46}, in terms of which the sc antiferromagnetic
structure is described by the state
\begin{equation}\label{Eq7}
(\alpha X_5+\beta Y_5+\gamma Z_5) ,\quad \alpha^2+\beta^2+\gamma^2=1
\end{equation}
with the energy per spin proportional to \cite{Lutt46}
\begin{equation}\label{Eq8}
\alpha^2+\beta^2+\gamma^2 .
\end{equation}
According to the definition of $X_5$, $Y_5$, and $Z_5$, it implies the
three-dimensional (3D) degeneracy of this antiferromagnetic (AF) state
shown in Fig.~\ref{Fig1} (a), with local moments, in units of $|\Mb|$, of
the form \cite{Lutt46}
\begin{eqnarray}\label{Eq9}
&&\lefteqn{\ab=(\alpha,\beta,\gamma)
,}\hspace{10em} \bb=(\alpha,-\beta,-\gamma) ,\nonumber\\
&&\lefteqn{\cb=(-\alpha,\beta,-\gamma) ,}\hspace{10em}
\db=(-\alpha,-\beta,\gamma) .
\end{eqnarray}

\begin{figure}[t]
\resizebox{0.95\hsize}{!}{\includegraphics{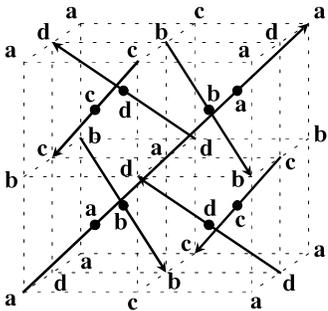}}
\caption{The structure of interpenetrating ferromagnetic chains shown by
the solid arrows in the nearest-neighbor directions as a particular case
of Fig.~\protect{\ref{Fig1}} (b1) at $\ab=(\protect{\textstyle%
\frac{1}{\protect\sqrt{3}},\frac{1}{\protect\sqrt{3}},
\frac{1}{\protect\sqrt{3}}})$. The first sublattice is depicted by the
dotted lines, whereas the filled circles exhibit the second sublattice.}
\label{Fig2}
\end{figure}

The basic AF bcc structure is unstable, in agreement with Table
\ref{Table1}. Nevertheless, it can be specified in the same manner. If
the two sc sublattices are described by the states
\begin{equation}\label{Eq10}
(\alpha X_7+\beta Y_7+\gamma Z_7) ,\quad
(\xi X_6+\eta Y_6+\zeta Z_6)
\end{equation}
normalized like (\ref{Eq7}), then the energy of interaction of these
sublattices is specified by \cite{Lutt46}
\begin{equation}\label{Eq11}
\max\{\alpha\eta+\beta\zeta+\gamma\xi\}\to\quad
\xi=\gamma,\quad \eta=\alpha,\quad \zeta=\beta,
\end{equation}
for the expression in the curly braces can be treated as a scalar product
of two unit vectors which must be parallel. It implies the 3D
configuration shown in Fig.~\ref{Fig1} (b1), where apart from (\ref{Eq9}),
the local spin states defined as
\begin{eqnarray}\label{Eq12}
&&\lefteqn{\eb=(\gamma,\alpha,\beta) ,}\hspace{10em}
\fb=(\gamma,-\alpha,-\beta) ,\nonumber\\
&&\lefteqn{\gb=(-\gamma,\alpha,-\beta) ,}\hspace{10em}
\hb=(-\gamma,-\alpha,\beta)
\end{eqnarray}
take place. The interchange of the two sublattices leads to the twofold
degeneracy of such states, as shown in Table \ref{Table1}. In other words,
the two rotations of the lattice in Fig.~\ref{Fig1} (a) about the (1,1,1)
axis, but without rotating spins, generate the first sublattice here. It
is worth noting that the particular cases of $\ab$ along one of the four
cubic diagonals are associated with the idea of Sauer \cite{Saue40} about 
ferromagnetic chains along the nearest neighbors, as shown in 
Fig.~\ref{Fig2}, where the cubic chain packing appears to be similar to 
the chain cubic structure intrinsic in TlHF$_2$ at low temperatures 
\cite{Khol83}.

Another possibility is associated with the states
\begin{equation}\label{Eq13}
(\beta Y_7+\gamma Z_6) ,\quad (\gamma Y_7+\beta Z_6),
\end{equation}
which imply the configuration shown in Fig.~\ref{Fig1} (b2) at
\begin{equation}\label{Eq14}
\abt=(0,\beta,\gamma) ,\quad \nbt=(0,\gamma,\beta).
\end{equation}
The cyclic interchange of the parameters in (\ref{Eq14}) explains the
threefold degeneracy of this two-dimensional (2D) set, as indicated in
Table \ref{Table1}. Note that all the states above follow from the 
'nearest-neighbor' treatment of Sauer \cite{Saue40} as well and this set
of possibilities is complete.

The same 'nearest-neighbor' treatment enables us to extend the solution
typical of the bcc structure to the diamond one. Indeed, any sc lattice
can be regarded as a combination of two fcc sublattices, as it happens in
the NaCl structure, but with equal energy effects at body-centered
positions.  So, upon removing one of the fcc sublattices from
either of the sc sublattices in the bcc structure, a diamond structure
arises. As a result, the total set of solutions for the diamond structure
is still defined either by (\ref{Eq9}) and (\ref{Eq12}), or by
(\ref{Eq14}), as shown in Fig.~\ref{Fig1} (c1) and Fig.~\ref{Fig1} (c2),
respectively, with the corresponding degrees of degeneracy. The chain 
structure similar to that in Fig.~\ref{Fig2}, but with two alternating 
spin-spin distances along every chain, takes place here as well.

For completeness, we also consider the fcc lattice, where the situation
turns out to be much more complicated and dramatic. In terms of unit
vectors $\ab$, $\nb_1$, $\nb_2$, and $\nb_3$ specifying four sc
sublattices of the same form, both the above treatments predict the
limiting antiferromagnetic structure drawn in Fig.~\ref{Fig1} (d).
On introducing the unit vector components along the principal axes:
\begin{equation}\label{Eq15}
\ab=(\alpha,\beta,\gamma) ,\quad \nb_l=(\xi_l,\eta_l,\zeta_l) ,
\end{equation}
one can show that the lowest energy of such a structure corresponds to
the condition
\begin{eqnarray}\label{Eq16}
&&(\eta_1+\zeta_3-\alpha)^2+(\xi_1+\zeta_2-\beta)^2\nonumber\\
&&{}+(\xi_3+\eta_2-\gamma)^2+(\xi_2+\eta_3+\zeta_1)^2=0 ,
\end{eqnarray}
with the obvious solutions
\begin{eqnarray}\label{Eq17}
&&\eta_1+\zeta_3-\alpha=0 ,\qquad \xi_1+\zeta_2-\beta=0 ,\nonumber\\
&&\xi_3+\eta_2-\gamma=0 ,\qquad \xi_2+\eta_3+\zeta_1=0 .
\end{eqnarray}
Further it is expedient to write down the unit vector normalization in
the form
\begin{eqnarray}
&&\xi_l=A_l ,\qquad\eta_l=C_lP_l ,\qquad\zeta_l=C_lQ_l ,\label{Eq18}\\
&&A_l^2+C_l^2+1 ,\qquad P_l^2+Q_l^2=1 \label{Eq19}
\end{eqnarray}
corresponding to spherical coordinates. According to (\ref{Eq16}) and
(\ref{Eq19}), it is clear that apart from the reference unit vector 
$\ab$, which is assumed to be arbitrary, two additional parameters 
describing (\ref{Eq18}) remain variable as well. In particular, we 
choose $\xi_1\equiv A$ and $\xi_2\equiv B$ as such parameters here. 
On substituting (\ref{Eq18}) into (\ref{Eq17}) and resolving those 
relations with account of (\ref{Eq19}), one can show that the other 
components take the form
\begin{eqnarray}
&&\quad\zeta_2=\beta-A ,\label{Eq20}\\
&&\cases{{\displaystyle\eta_2=\pm\sqrt{1-(\beta-A)^2-B^2}}\cr
{\displaystyle\xi_3=\gamma\mp\sqrt{1-(\beta-A)^2-B^2}}} ,\label{Eq21}\\
&&\cases{{\displaystyle\eta_1=\frac{-\alpha(A^2-\xi_3^2-\alpha^2-B^2)\pm
B\mOmega}{2(\alpha^2+B^2)}} ,\cr
{\displaystyle\zeta_1=\frac{B(A^2-\xi_3^2-\alpha^2-B^2)\pm
\alpha\mOmega}{2(\alpha^2+B^2)}} ,\cr
{\displaystyle\eta_3=\frac{-B(A^2-\xi_3^2+\alpha^2+B^2)\mp
\alpha\mOmega}{2(\alpha^2+B^2)}} ,\cr
{\displaystyle\zeta_3=\frac{\alpha(A^2-\xi_3^2+\alpha^2+B^2)\mp
B\mOmega}{2(\alpha^2+B^2)}} ,}\label{Eq22}\\
&&\hspace{-2em}\mOmega={\textstyle\sqrt{4(1{-}A^2)(\alpha^2{+}B^2)
-(A^2{-}\xi_3^2{-}\alpha^2{-}B^2)^2}} ,\label{Eq23}
\end{eqnarray}
where the choice of an upper or lower sign is common to both the relations
(\ref{Eq21}) and the same is right for all four relations (\ref{Eq22}),
which case is independent of (\ref{Eq21}). Altogether, four two-parameter
sets of solutions as functions of $\alpha$, $\beta$, and $\gamma$ exist,
as pointed out in Table \ref{Table1}. The change in $A$ and $B$ is
restricted by the following inequalities:
\begin{eqnarray}
&&\hspace{-1em}|A|\leq1 ,\qquad \mp2\gamma-\gamma^2\leq(\beta-A)^2+B^2\leq1 ,
\label{Eq24}\\
&&\hspace{-1em}(A^2-\xi_3^2-\alpha^2-B^2)^2\leq4(1-A^2)(\alpha^2+B^2)
\label{Eq25}
\end{eqnarray}
which emerge from (\ref{Eq19}) and from the conditions that the square
roots in (\ref{Eq21}) and (\ref{Eq23}) be real, an upper or lower sign in
(\ref{Eq24}) agrees with that in (\ref{Eq21}).

In particular, if $A=\beta$ and $B=\lambda=\pm1$, then relations
(\ref{Eq24}) and (\ref{Eq25}) hold automatically and solutions
(\ref{Eq20})--(\ref{Eq23}) are reduced to
\begin{eqnarray}
&&\quad\eta_2=\zeta_2=0 ,\qquad \xi_3=\gamma ,\label{Eq26}\\
&&\cases{{\displaystyle\eta_1=\frac{\alpha(1-\beta^2)\pm\lambda
\sqrt{(1-\beta^2)(1-\gamma^2)}}{1+\alpha^2}} ,\cr
{\displaystyle\zeta_1=\frac{\lambda(\beta^2-1)\pm\alpha
\sqrt{(1-\beta^2)(1-\gamma^2)}}{1+\alpha^2}} ,\cr
{\displaystyle\eta_3=\frac{\lambda(\gamma^2-1)\mp
\alpha\sqrt{(1-\beta^2)(1-\gamma^2)}}{1+\alpha^2}} ,\cr
{\displaystyle\zeta_3=\frac{\alpha(1-\gamma^2)\mp\lambda
\sqrt{(1-\beta^2)(1-\gamma^2)}}{1+\alpha^2}} .}\label{Eq27}
\end{eqnarray}
The existence of four branches of solutions (\ref{Eq20})--(\ref{Eq23})
ensues from (\ref{Eq26})--(\ref{Eq27}) by continuity.

The case of $\alpha=B=0$ implies $\abt=(0,\beta,\gamma)$ and
$\nbt_2=(0,\eta_2,\zeta_2)$. It is degenerate in the representation at
hand and can be studied directly from (\ref{Eq17}) giving rise to
\begin{equation}\label{Eq28}
\eta_3=-\zeta_1 ,\qquad\zeta_3=-\eta_1 ,\qquad
\eta_1^2+\zeta_1^2=1-\xi_1^2 .
\end{equation}
According to (\ref{Eq19}), there are two solutions for the other 
components here. The first one is defined by
\begin{equation}\label{Eq29}
\xi_1=\xi_3=0 ,\qquad\eta_2=\gamma ,\qquad\zeta_2=\beta ,
\end{equation}
in which case the consistent rotation of $\abt$ and $\nbt_2$ in the plane
normal to the first axis is accompanied by an independent consistent
rotation of the couple of $\nb_1$ and $\nb_3$ in the same plane. In other
words, (\ref{Eq28}) and (\ref{Eq29}) generate a one-parameter continuous
solution. The second solution in question is of the form
\begin{equation}\label{Eq30}
\cases{\lefteqn{\xi_1=-\xi_3=\beta-\gamma ,}\hspace{9em}\nbt_2=\abt&
at $\beta\gamma>0$,\cr
\lefteqn{\xi_1=\xi_3=\beta+\gamma ,}\hspace{9em}\nbt_2=-\abt&
at $\beta\gamma<0$.}
\end{equation}
According to (\ref{Eq28}) and (\ref{Eq30}), one can see that an
independent consistent precession of $\nb_1$ and $\nb_3$ about the first
axis is admissible, while both $\beta$ and $\gamma$ are nonzero and so
determine the angle of precession. In this case the rotations of both
$\abt$ and $\nbt_2$ are connected. But if either $\beta$ or $\gamma$ goes
through zero, then $\nbt_2$ undergoes the sudden inversion. At those
moments the precession of $\nb_1$ and $\nb_3$ disappears: both $\nb_1$ and
$\nb_3$ are along the first axis and one of them is inverted in a
jump-like manner as well.

Note that the solutions of similar types are to be expected at
$\beta=\eta_3=0$ and $\gamma=\zeta_1=0$. This is the reason that all three
pairs of these one-parameter two-dimensional solutions are pointed out in
Table \ref{Table1}, though the latter solutions, with including piecewise 
continuous ones, are contained in the general case of
(\ref{Eq20})--(\ref{Eq23}).

In summary, both the 'nearest-neighbor' and group-theoretic approaches are
suitable for investigating the lowest energy antiferromagnetic
configurations. The latter one appears to be more effective upon
describing the sc and bcc lattices, but the former is favorable for
extending the foregoing results to the diamond case. Of course, the
spherical degeneracy discussed is the most important for antiferromagnetic
ground states in the sc and diamond lattices, where it accounts for the
high isotropic susceptibility recorded experimentally \cite{Whit93}. On
the other hand, the spherical degeneracy revealed is a general property of
the cubic symmetry. Thus the peculiar features of the foregoing results
relevant to the sc, bcc, and fcc lattices are to be important for 
understanding the energy splitting upon lowering the lattice symmetry.

\end{document}